\documentclass[14pt]{article}
\usepackage{srcltx}
\usepackage[T2A]{fontenc}
\usepackage[utf8]{inputenc}
\usepackage[english]{babel}
\usepackage{graphicx}
\usepackage{subcaption}
\usepackage{amsmath,amsfonts,amssymb,amsthm,mathtools} 
\usepackage{booktabs}
\usepackage{xcolor}

\begin{document}

\title{ Stellar Critical Parameters in the Uniform Density Approximation}

\author{%
G.\,S.\,Bisnovatyi-Kogan\thanks{Space Research Institute, Russian Academy of Sciences, Moscow, Russia; National Research Nuclear University MEPhI (Moscow Engineering Physics Institute), Moscow, Russia; Moscow Institute of Physics and Technology, Dolgoprudny, Moscow oblast, Russia},
E.\,A.\,Patraman\thanks{Space Research Institute, Russian Academy of Sciences, Moscow, Russia; Moscow Institute of Physics and Technology, Dolgoprudny, Moscow oblast, Russia}
}

\date{}
 
\maketitle

\begin{abstract}
      Stellar models are calculated in the approximation of a uniform density distribution. Variational method was used for determination of the boundary of a stability loss, for stellar masses in the range from 2 up to $10^5$ $M_{\odot}$.
       The effects of the general relativity had been taken into account. The equation of state in the temperature and density ranges 
       $10^9< T < 10^{10} K$,  $10^5< \rho < 10^{10} \frac{\text{g}}{\text{сm}^3}$ had been taken from the work of Imshennik and Nadyozhin (1965). The critical parameters for the values of entropy and stellar masses differ from more accurate values, obtained using a more complicated variant of accepted density distribution, not more than 12$~\%$. 
\end{abstract}

\bigskip


\section{Introduction.}

When studying the structure of white dwarfs, it was
discovered that their equilibrium is possible only for
masses not exceeding a certain limit, which is known
as the Chandrasekhar limit. For the oxygen-carbon
chemical composition, this limit is equal to
 $\approx 1.46\, M_\odot$.  The first conclusion about the existence of an upper mass limit for cold stars, the equilibrium of which is maintained by the pressure of relativistically degenerate electrons, was made in the work of Stoner \cite{stoner}, who
considered the model of a white dwarf of uniform density. More precise results for the maximum mass of a
white dwarf were obtained in \cite{chandra, landau_pred}. 

Stars with masses less than the Chandrasekhar
limit cool over time and become cold white dwarfs
with zero entropy S = 0 \cite{zeld1}. Stars with greater mass in the process of cooling and loss of entropy, they lose
stability when $S \neq 0$, after which they collapse, turning into a neutron star or a black hole. 

The reasons for the loss of stability of stars of different masses are different. For stars with mass $ 1.2M_{\odot}<M< 3M_{\odot}$ the main effect is introduced by
the neutronization of matter and the effects of the general theory of relativity (GTR) \cite{zeld1,kaplan, bkwd}. As the mass of the star increases, up to $\approx 500 M_{\odot}$, the cause of instability is the dissociation of iron, which, when $M \approx 500 - 5000 M_{\odot}$ gives way to the pair production \cite{  bk}. At $M > 5\cdot10^3 M_{\odot}$, the main role in the loss of stability  of a star is played by the effects of general relativity \cite{fowler, zeld2}.

  In this paper, a uniform density distribution is
adopted to calculate the critical parameters of stars in
the mass range $2M_{\odot} < M < 10^5M_{\odot}$.The equation of state taking into account the pair production and dissociation of iron was taken from \cite{ bk, imsh}.  Small corrections to GR for stars in the uniform density approximation were calculated in \cite{bkp}. 

For an approximate consideration of homogeneous
stars, the variational method was used, in which the
star was assumed to be isentropic. The results of the
work, presented in the figures and tables, determine
the dependence of the critical density $\rho_{cr}(M)$ and
entropy  $S_{cr}(M)$ from the mass, in the mass range from $2 M_{\odot}$ to supermassive stars. Figure $S_{cr}(M)$ in this
region of masses differs from the model from work \cite{bk} by no more than 12$\%$. 

\section{Variational method}

To calculate the critical parameters of stars, a variational method was used, in which the density distribution is assumed to be uniform, and this density is the
only parameter, when changing which, changes in the
star occur homologically.

The total energy of a homogeneous star, taking into
account the small effects of general relativity, is written
in the form \cite{ bkp, landau}
\begin{equation}
\label{eq1}
    \varepsilon = E_T M - \frac{3}{5}G\big( \frac{4\pi}{3}\big)^{1/3}M^{5/3}\rho^{1/3} - 1.982\frac{G^2}{c^2}M^{7/3}\rho^{2/3}
\end{equation}
Here, $E_T$ is the internal energy of a unit mass, the second term is the gravitational energy of a homogeneous
sphere, the third term is the correction for general relativity for a star of uniform density. Equilibrium state
and stability condition for a star of uniform density
with mass $M$ are written in the form:
\begin{equation}
    3MP(\rho, T)\rho^{-4/3} - \frac{3}{5}G\big( \frac{4\pi}{3}\big)^{1/3}M^{5/3} - 3.964\frac{G^2}{c^2}M^{7/3}\rho^{1/3} = 0
\label{eq:equl}
\end{equation}
\begin{equation}
    9MP(\rho, T)\rho^{-5/3}(\gamma - 4/3) - 3.964\frac{G^2}{c^2}M^{7/3} > 0
    \label{eq:equla}
\end{equation}
The star is considered isentropic, therefore, to solve
the system (2),(3), knowledge of the functions is
required $P(\rho), \gamma(\rho)$ along isentropes over a wide range of densities and temperatures.

\section{Equation of state}

When considering the stability of a star, it was
assumed that the star had completely exhausted its
nuclear sources, and its chemical composition was
determined by iron. At high temperatures and
densities, $\alpha$ - particles, protons $p$ and neutrons $n$ in thermodynamic equilibrium appears. Equation of state in a range of densities $1<\rho_5< 10^{5} $ and temperatures $1 < T_9 < 20 $, where $\rho_5 = \rho/10^5 {g}/{cm}^{3}$, $T_9 = {T}/{10^9} K$, calculated in work \cite{imsh}, where the relativistic degeneracy of electrons and positrons, the production of pairs
and photons, $\beta -$processes, at
nuclear equilibrium in this mixture with $^{56}Fe$,  including the first seven excited levels, $^4He$,  $p, n$, were taken into account. Function $P(\rho,\, T)$  in the square of the densities 1<$\rho_5$< $10^{5}$ and temperatures $1 < T_9 < 20 $, , calculated in the work \cite{imsh} was used by us in calculating the dependencies $P(\rho)$ for
11 isentropes, which are presented in the Tables \ref{table:nad1},  \ref{table:nad2}. The isentrope values in the tables are given in units of
[erg/K/g], since these units of measurement were
chosen in the work \cite{imsh}. To go to dimensionless
entropy per nucleon, these values must be divided by $k/m_p \approx 8.31\cdot10^7$ erg/g/K.
 \\

\begin{table}[]
\caption{The dependence $P(\rho)$,  obtained on the basis of the results of work \cite{imsh} and by solving the system of equations (4) - (10) for isentropes: $S_{1} = 0.01$, $S_{2} = 0.01585 $ , $S_{3} = 0.02512 $ , $S_{4} = 0.03981 $ , $S_{5} = 0.0631 $ , $S_{6} = 0.1 $ .  Isentrope values are given in units $10^{10}$ erg/(K g).  The indices of the pressures correspond to the isentropic number..}
\label{table:nad1}
\centering
\begin{tabular}{|l|l|l|l|l|l|l|}
\hline
$lg(\rho/10^7)$ & $lg(P_1/10^{25})$ & $lg(P_2/10^{25})$ & $lg(P_3/10^{25})$ & $lg(P_{4}/10^{25})$ & $lg(P_{5}/10^{25})$ & $lg(P_{6}/10^{25})$\\ \hline

-5.6 &           &           &           &           &           & -7.2648 \\\hline
-5.4 &           &           &           &           & -7.3823 & -6.9913 \\\hline
-5.2 &           &           &           &           & -7.1026 & -6.7179 \\\hline
-5.0 &           &           &           &           & -6.8232 &-6.4446 \\\hline
-4.8 &           &           &           & -7.0077 & -6.5444 & -6.1715 \\\hline
-4.6 &           &           &           & -6.7155 & -6.2657 & -5.8986 \\\hline
-4.4 &           &           &           & -6.4246 & -5.9875 & -5.6258 \\\hline
-4.2 &           &           &           & -6.1347 & -5.7097 & -5.3531 \\\hline
-4.0 &           &           &           & -5.8460 & -5.4323 & -5.0806 \\\hline
-3.8 &           &           & -6.0810 & -5.5584 & -5.1552 & -4.8085 \\\hline
-3.6 &           &           & -5.7687 & -5.2719 & -4.8786 & -4.5365 \\\hline
-3.4 &           &           & -5.4596 & -4.9863 & -4.6024 & -4.2646 \\\hline
-3.2 &           &           & -5.1536 & -4.7018 & -4.3267 & -3.9932 \\\hline
-3.0 &           & -5.3468 & -4.8506 & -4.4182 & -4.0513 & -3.7231 \\\hline
-2.8 &           & -5.0217 & -4.5503 & -4.1356 & -3.7767 & -3.4565 \\\hline
-2.6 &           & -4.6990 & -4.2526 & -3.8541 & -3.5070   & -3.1993   \\\hline
-2.4 &           & -4.3793 & -3.9585 & -3.5743   & -3.2373   & -2.9420   \\\hline
-2.2 &           & -4.0628 & -3.6643 & -3.2945 & -2.9676 & -2.6848 \\\hline
-2.0 & -5.1729 & -3.7258 & -3.3828   & -3.0100   & -2.7113   & -2.4356 \\\hline
-1.8 & -4.5969 & -3.4194 & -3.1109 & -2.7209 & -2.4638 & -2.1849   \\\hline
-1.6 & -4.0650 & -3.1161 & -2.7779 & -2.4676 & -2.2095   & -1.9341   \\\hline
-1.4 & -3.5737 & -2.8157 & -2.4768 & -2.2013 & -1.9553 & -1.6834 \\\hline
-1.2 & -3.1196 & -2.5183 & -2.1982 & -1.9297 & -1.6739 & -1.3964 \\\hline
-1.0 & -2.6993 & -2.2240 & -1.9329 & -1.6607 & -1.3954 & -1.1155 \\\hline
-0.8 & -2.3093 & -1.9326 & -1.6717 & -1.3988 & -1.1304 & -0.8516 \\\hline
-0.6 & -1.9462 & -1.6443 & -1.4052 & -1.1366 & -0.8682 & -0.5998 \\\hline
-0.4 & -1.6066 & -1.3589 & -1.1247 & -0.8637 & -0.5942 & -0.3518 \\\hline
-0.2 & -1.2869 & -1.0766 & -0.8334 & -0.5834 & -0.3168 & -0.1055 \\\hline
0.0  & -0.9839 & -0.7973 & -0.5465 & -0.3116 & -0.0672 & 0.1345  \\\hline
0.2  & -0.6940 & -0.5210 & -0.2767 & -0.0617 & 0.1345  & 0.3649  \\\hline
0.4  & -0.4138 & -0.2481 & -0.0248 & 0.1662  & 0.3180  & 0.5882  \\\hline
0.6  & -0.1398 & 0.0213  & 0.2115  & 0.3759  & 0.5236  & 0.8079  \\\hline
0.8  & 0.1313  & 0.2860  & 0.4381  & 0.5763  & 0.7596  & 1.0265  \\\hline
1.0  & 0.4030  & 0.5442  & 0.6642  & 0.7820  & 1.0018  & 1.2450  \\\hline
1.2  & 0.6774  & 0.7951  & 0.8974  & 1.0041  & 1.2306  & 1.4645  \\\hline
1.4  & 0.9511  & 1.0426  & 1.1376  & 1.2396  & 1.4508  & 1.6863  \\\hline
1.6  & 1.2195  & 1.2913  & 1.3824  & 1.4820  & 1.6731  & 1.9120  \\\hline
1.8  & 1.4816  &1.5421  & 1.6288  & 1.7254  & 1.9005   & 2.1432  \\\hline
2.0  & 1.7400  & 1.7922  & 1.8725  & 1.9648  & 2.1289  & 2.3813  \\\hline
2.2  & 1.9968  & 2.0389  & 2.1105  & 2.1964  & 2.3549  & 2.6278  \\\hline
2.4  & 2.2525  & 2.2830  & 2.3442  & 2.4229  & 2.5801  & 2.8844  \\\hline
2.6  & 2.5068  & 2.5256  & 2.5758  & 2.6481  & 2.8074  & 3.1524  \\\hline
2.8  & 2.7596  & 2.7679  & 2.8080  & 2.8760  & 3.0396  & 3.4335  \\\hline
3.0  & 3.0106  & 3.0112  & 3.0431  & 3.1105 & 3.2795  & 3.7290  \\\hline

\end{tabular}
\end{table}

\begin{table}[]
\caption{The dependence $P(\rho)$, obtained on the basis of the results of work \cite{imsh} and by solving the system of equations (4) - (10) for isentropes: $S_{7} = 0.1585$, $S_{8} = 0.2512 $ , $S_{9} = 0.3981 $ , $S_{10} = 0.631 $ , $S_{11} = 1 $.  Isentrope values are given in units $10^{10}$ erg/(K g). The indices of the pressures correspond to the isentropic number.}
\label{table:nad2}
\centering
\begin{tabular}{|l|l|l|l|l|l|}
\hline
$lg(\rho/10^7)$ & $lg(P_7/10^{25})$ & $lg(P_8/10^{25})$ & $lg(P_9/10^{25})$ & $lg(P_{10}/10^{25})$ & $lg(P_{11}/10^{25})$  \\ \hline

-6.8 &           &           &           &   & -7.3598 \\ \hline
-6.6 &           &           &     & -7.3721 & -7.0926  \\ \hline
-6.4 &           &       & -7.3908 & -7.1046 & -6.8255   \\ \hline
-6.2 &         & -7.4196 & -7.1228 & -6.8371 & -6.5583   \\ \hline
-6.0 & -7.4651 & -7.1507 & -6.8547 & -6.5696 & -6.2911   \\ \hline
-5.8 & -7.1946 & -6.8817 & -6.5867 & -6.3021 & -6.0239    \\ \hline
-5.6 & -6.9240 & -6.6128 & -6.3187 & -6.0347 & -5.7568   \\\hline
-5.4 & -6.6535 & -6.3439 & -6.0507 & -5.7672 & -5.4896  \\\hline
-5.2 & -6.3830 & -6.0750 & -5.7828 & -5.4998 & -5.2225 \\\hline
-5.0 & -6.1126 & -5.8063 & -5.5148 & -5.2323 & -4.9553  \\\hline
-4.8 & -5.8423 & -5.5375 & -5.2469 & -4.9649 & -4.6882 \\\hline
-4.6 & -5.5722 & -5.2688 & -4.9790 & -4.6975 & -4.4214   \\\hline
-4.4 & -5.3021 & -5.0002 & -4.7111 & -4.4303 & -4.1566    \\\hline
-4.2 & -5.0321 & -4.7316 & -4.4434 & -4.1642 & -3.8978    \\\hline
-4.0 & -4.7623 & -4.4631 & -4.1763 & -3.9027 & -3.6465   \\\hline
-3.8 & -4.4924 & -4.1949 & -3.9124 & -3.6496 & -3.3998 \\\hline
-3.6 & -4.2228 & -3.9284 & -3.6552 & -3.4017 & -3.1552  \\\hline
-3.4 & -3.9540 & -3.6669 & -3.4059 & -3.1566 & -2.9100   \\\hline
-3.2 & -3.6881 & -3.4131 & -3.1594 & -2.9112 & -2.6622  \\\hline
-3.0 & -3.4282 & -3.1650 & -2.9134 & -2.6633 & -2.4108  \\\hline
-2.8 & -3.1747 & -2.9173 & -2.6653 & -2.4119 & -2.1557\\\hline
-2.6 & -2.9236 & -2.6653 & -2.4138 & -2.1567 & -1.8969  \\\hline
-2.4 & -2.6726 & -2.4133 & -2.1567 & -1.8963 & -1.6335    \\\hline
-2.2 & -2.4216 & -2.1612 & -1.8996 & -1.6358 & -1.3702     \\\hline
-2.0 & -2.1469 & -1.8833 & -1.6048 & -1.3363 & -1.0879     \\\hline
-1.8 & -1.8861 & -1.6242 & -1.3570 & -1.0803 & -0.8356       \\\hline
-1.6 & -1.6252 & -1.3651 & -1.1093 & -0.8243 & -0.5833    \\\hline
-1.4 & -1.3644 & -1.1060 & -0.8528 & -0.5771 & -0.3223     \\\hline
-1.2 & -1.0882 & -0.8337 & -0.5945 & -0.3278 & -0.0576     \\\hline
-1.0 & -0.8246 & -0.5754 & -0.3374 & -0.0705 & 0.2092      \\\hline
-0.8 & -0.5809 & -0.3374 & -0.0836 & 0.1978  & 0.4767      \\\hline
-0.6 & -0.3453 & -0.1027 & 0.1678  & 0.4681  & 0.7436      \\\hline
-0.4 & -0.1016 & 0.1507  & 0.4185  & 0.7289  & 1.0091      \\\hline
-0.2 & 0.1510  & 0.4202  & 0.6706  & 0.9784  & 1.2721  \\\hline
0.0  & 0.3983  & 0.6792  & 0.9268  & 1.2249  & 1.5315      \\\hline
0.2  & 0.6292  & 0.9082  & 1.1896  & 1.4769  & 1.7863      \\\hline
0.4  & 0.8473  & 1.1223  & 1.4615  & 1.7429  & 2.0356      \\\hline
0.6  & 1.0601  & 1.3450  & 1.7452  & 2.0316  & 2.2782      \\\hline
0.8  & 1.2764  & 1.5999  & 2.0432  & 2.3516  & 2.5132      \\\hline
1.0  & 1.5062  & 1.9105  & 2.3581  & 2.7114  & 2.7396      \\\hline
1.2  & 1.7601  & 2.3003  & 2.6925  & 3.1196  & 2.9563      \\\hline
1.4  & 2.0481  & 2.7929  & 3.0489  & 3.5847  & 3.1624      \\\hline
1.6  & 2.3807  & 3.4119  & 3.4299  & 4.1155  & 3.3567    \\\hline
1.8  & 2.7681  & 4.1807  & 3.8381  & 4.7204  & 3.5383     \\\hline
2.0  & 3.2206  & 5.1228  & 4.2761  & 5.4080  & 3.7061     \\\hline
\end{tabular}
\end{table}
 At lower temperatures and densities, the dissociation of iron is suppressed,  $\alpha,\, p,\, n$ do not appear. The
  only element remaining is iron, and the chemical
composition is considered homogeneous. The calculation of the equation of state and thermodynamic
functions along isentropes is carried out using the following analytical formulas \cite{bk} 
\begin{align}
\rho &= \frac{A m_p}{Q \pi^2} \left(\frac{k T}{c \hbar}\right)^3 (I_{1-} - I_{1+}), \label{eq:rho} \\
P &= \frac{\rho k T}{A m_p} + \frac{1}{3} \sigma T^4 + \frac{1}{3 \pi^2} \left(\frac{k T}{c \hbar}\right)^3 k T (I_{2-} + I_{2+}), \label{eq:P}\\
E &= \frac{3}{2} \frac{k T}{A m_p} + \frac{\sigma T^4}{\rho} + \frac{1}{\pi^2} \left(\frac{k T}{c \hbar}\right)^3 k T (I_{3-} + I_{3+}), \\
S &= \frac{k}{A m_p} \left\{ \frac{5}{2} + \ln \left[ \left( \frac{A m_p k T}{2 \pi \hbar^2} \right)^{3/2} \frac{g A m_p}{\rho} \right] \right\} + \frac{4 \sigma T^3}{3 \rho} + \label{eq:S} \\ 
&\quad  + \frac{k}{3 \pi^2 \rho} \left( \frac{k T}{c \hbar} \right)^3 \left[ 3(I_{3+} + I_{3-}) - 3b (I_{1-} - I_{1+}) + I_{2-} + I_{2+} \right],  \nonumber
\end{align}
where

\begin{align} 
I_{1\pm} &= \int_0^\infty \frac{x^2 \, dx}{1 + \exp(\sqrt{x^2 + \alpha^2} \pm b)}, \label{eq:I1} \\
I_{2\pm} &= \int_0^\infty \frac{x^4 \, dx}{\sqrt{\alpha^2 + x^2} \left[ 1 + \exp(\sqrt{x^2 + \alpha^2} \pm b) \right]}, \label{eq:I2}\\
I_{3\pm} &= \int_0^\infty \frac{\sqrt{x^2 + \alpha^2} \, x^2 \, dx}{1 + \exp(\sqrt{x^2 + \alpha^2} \pm b)} \label{eq:I3}.
\end{align}

Here, $b = \mu/kT$, $\alpha = m_e c^2/kT$, $\mu$ $-$ the chemical
potential of electrons, $m_p$ is the mass of the proton, the
mass of the nucleus was taken to be equal to $Am_p$, $m_e$ is the mass of an electron, $k$ $-$ Boltzmann constant, $\hbar$ $-$ Planck’s constant divided by $2\pi$, $\sigma = \pi^2 k^4 / 15c^3\hbar^3$ is the radiation density constant, $g$ is the statistical weight of the nucleus in the ground state. For iron $A = 56, Q = 26, g = 1; E$  is the energy of a unit of
mass, including the rest energy of electrons and positrons. 

Equation \eqref{eq:rho} is used to find $\mu(\rho, T)$. This equation was solved numerically by Newton’s method with relative accuracy $10^{-8}$. Integrals (\ref{eq:I1}) - (\ref{eq:I3}) were calculated
using a method based on the generalized Gaussian
numerical integration method. On its basis, in works \cite{dima,blinnikov} approximate expressions were obtained for integrals describing the thermodynamic properties of
the electron–positron gas, in the form of interpolation formulas for functions of two variables depending on
temperature and chemical potential

\begin{equation}
n_- = \frac{1}{\lambda^3\alpha^3} \cdot G_1(\alpha, \chi),
\end{equation}
\begin{equation}
P_- = \frac{mc^2}{3\lambda^3} \cdot \frac{1}{\alpha^4} G_2(\alpha, \chi),
\end{equation}

\begin{equation}
E_- = \frac{mc^2}{\lambda^3 \rho} \cdot \frac{1}{\alpha^4} G_3(\alpha, \chi) - mc^2 N_A Y,
\end{equation}

\begin{equation}
S_- = \frac{k}{3\lambda^3 \rho}\cdot \frac{1}{\alpha^4} \left[ 3G_3 - 3\alpha w G_1 + G_2 \right],
\end{equation}

\begin{equation}
G_1(\alpha, \chi) = \int_0^\infty \frac{(u + \alpha) \sqrt{u(u + 2\alpha)}}{1 + \exp(u - \chi)} \, du,
\end{equation}

\begin{equation}
G_2(\alpha, \chi) = \int_0^\infty \frac{u(u + 2\alpha) \sqrt{u(u + 2\alpha)}}{1 + \exp(u - \chi)} \, du,
\end{equation}

\begin{equation}
G_3(\alpha, \chi) = \int_0^\infty \frac{(u + \alpha)^2 \sqrt{u(u + 2\alpha)}}{1 + \exp(u - \chi)} \, du,
\end{equation}

\begin{equation}
 \alpha=\frac{m_e c^2}{kT}, \quad
 \chi = \frac{\mu - m_ec^2}{kT}, \quad \omega  = \frac{\mu}{m_ec^2}, \quad \lambda^3 = \pi^2\Big(\frac{\hbar}{m_ec^2}\Big)^3
\end{equation}
The expressions for the positron gas can be obtained by
replacing  $\chi$ on $-\chi-2\alpha$. Integrals $G_1$, $G_2$, $G_3$ were calculated using the series expansion method used
in  \cite{blinnikov}. 

For a highly degenerate gas, the error in calculating
the entropy when expanding into series increases, so it
was calculated using asymptotic formulas  \cite{bkbook}.

\begin{equation}
    S_{-} = \frac{m_e^2c}{3\hbar^3\rho}k^2Ty\sqrt{y^2 + 1}, \quad \text{where } \quad y = \Big(\frac{1.027\rho}{10^6\mu_z} \Big)^{1/3}
\end{equation}

In \cite{bk}, a similar version of the Gauss method \cite{krylov} was used to calculate integrals \eqref{eq:I1} - \eqref{eq:I3}, which, when $\chi \leq 0.6$ gives a discrepancy in the calculation of integrals compared to the method from  \cite{blinnikov} of no more
than 0.2$\%$. 

After calculating the thermodynamic functions \eqref{eq:P}, \eqref{eq:S}  interpolation was performed using the spline
method to find the values $\rho, T, P$, along the isentropes from the Tables 1, 2. Functions $P(\rho)$ along these
isentropes were smoothly stitched with the corresponding isentropes $P(\rho)$, obtained from work \cite{imsh}, and were used to solve equation \eqref{eq:equl}. 

\section{Results}
Equation \eqref{eq:equl} defines the analytical dependence $M(\rho)$  for a star in the uniform density approximation \cite{bkp}:
\begin{equation}
\label{eq:M_rho}
    M(\rho) = \Big(\frac{-\frac{3}{5}G\big(\frac{4\pi}{3} \big)^{1/3} + \sqrt{(\frac{3}{5}G\big(\frac{4\pi}{3} \big)^{1/3})^2 + 47.568\frac{G^2}{c^2}\rho^{-1}P}}{7.928\frac{G^2}{c^2}\rho^{1/3}}\Big)^{3/2}
\end{equation}
From formula \eqref{eq:M_rho} for a series of isentropes, equilibrium curves M($\rho$) were obtained (see Fig. 1). The
maxima of the curves correspond to the critical state
when the star loses its stability \cite{kosmog}.
We used this
method of finding critical parameters instead of solving equation
  \eqref{eq:equla}, which would be a more cumbersome
procedure.

In Table \ref{table:critical} critical values 
 $S_{u, cr}, M_{cr}/M_{\odot}$, 
$\rho_{u, cr}$ for a homogeneous model are given. 
In Table \ref{table:ratio}  a comparison of the critical parameters $-$ density and entropy of
a homogeneous model is given ($\rho_{u, cr},\, S_{u, cr}$) with the
corresponding values of central density and the same
entropy ($\rho_{c, cr}\, S_{c, cr}$) models from work \cite{bk}.
 In this work,
the same variational method was used to find the critical parameters, but for density profiling, the solution
of the Lane$-$Emden equation \cite{chandrasekhar} with the polytropic index n = 3 was used .
In this case, the integrals used were calculated numerically, in contrast to the
homogeneous density used here, which allows for the
analytical calculation of all integrals, which significantly simplifies the consideration.

Dependencies $\rho_{cr}(M)$ and $S_{cr}(M)$ in the case of a
uniform density distribution, they were constructed
based on interpolation of data from Table \ref{table:critical} , they are
shown in Figs. 2 and 3, respectively. In addition to
this, Figs. 2 and 3 show the dependencies $\rho_{cr}(M)$ and $S_{cr}(M)$ for the model from work \cite{bk}, while the value $\rho_{cr}$  corresponds to the central density of a star in a critical state.

Comparing the critical parameters of homogeneous stars with the critical parameters of stars in the
isentropic model from \cite{bk}, we can conclude that the
 critical entropy values for the same mass in the homogeneous model differ from the same values in the
model from \cite{bk}
  by no more than 12 $\%$, and the density
ratios $\rho_{c,cr}/\rho_{u,cr} $ are within the range of 2.15 - 11. 

This is due to the fact that in the work \cite{bk}, from
which the upper curve in these figures was taken, $\rho_{cr}(M)$ corresponds to the central density of a star with
a density distribution similar to that of a polytropic
model with $\gamma=4/3$, and the lower one corresponds to
the homogeneous model of this work. In this regard,
the difference in critical densities in these two models
refers to different quantities that differ significantly.
The only identical parameter in both cases is entropy,
so it is correct to compare the dependencies $S_{cr}(M)$.

In the polytropic model with $n=3$, used in the
work \cite{bk},  the central density is $\approx$ 54 times greater than
the average density of the star. Using this average density instead of the central one when plotting the equilibrium models in Fig. 2, we obtain a much larger deviation from the homogeneous model. Instead of the
above-mentioned excess of the central density over the
density of the homogeneous model for the same mass,
we get that the average density according to the polytropic model turns out to be 4.9$-$25.1 times less than
the uniform density for the same mass. 

\begin{table}[!ht]
    \centering
    \caption{Dependence of critical density $\rho_{u,cr}$ and entropy $S_{u, cr}$ from the mass of the star in homogeneous models.}
    \label{table:critical}
    \begin{tabular}{|l|l|l|l|l|}
    \hline
        $S_{u, cr} / 10^{10} erg/(g \cdot K)$ & $\frac{M_{cr}}{M_{\odot}}$ & $\rho_{u,cr}/10^5, g/cm^3 $ \\ \hline
        
        0.01 & 2.06 & 2650  \\ \hline
        0.01585 & 3.37 &354\\ \hline
        0.0205 &  5.34   &  237   \\ \hline    
        0.02512 & 7.3 & 120 \\ \hline
        0.0316 & 11.46  & 96.9  \\ \hline
        0.03981 & 16.74 & 67.6  \\ \hline
        0.0631 & 41.87 & 43.7\\ \hline
        0.0682 &  50.5& 38.9  \\ \hline
        0.099 & 102.61 & 10.3 \\ \hline
        0.1 & 104.3 & 9.33 \\ \hline
        0.1585 & 289.9 & 5.37 \\ \hline
        0.213 & 520& 4.9  \\ \hline
        0.2512 & 681.3 & 4.57  \\ \hline
        0.276 & 1200 & 0.015 \\ \hline
        0.3981 & 2690 & 0.0093 \\ \hline
        0.631 & 7024 & 0.0025 \\ \hline
        1 & 18135 & 1.0889e-04 \\ \hline
        
    \end{tabular}
\end{table}

\begin{table}[!ht]
\centering
    \caption{Comparison of critical parameters: central density and entropy from the model of work \cite{bk} ($\rho_{c, cr},\, S_{c, cr}$) with
the density and entropy of a homogeneous model
 ($\rho_{u, cr},\,S_{u, cr}$) with the same mass}
    \label{table:ratio}
\begin{tabular}{|l|l|l|}
\hline
    $M/M_{\odot}$ & $\frac{\rho_{c, cr}}{\rho_{u, cr}}$ &  $\frac{S_{c, cr}}{S_{u, cr}}$  \\ \hline
    3   &   2.85  &    1.12      \\ \hline
    5   &   4  &   1.06   \\ \hline
    10   &   4  &  1.07   \\ \hline
    50   &   2.65  &  1.006        \\ \hline
    100  &  3.92   &   1.01   \\ \hline
    500  &  11.3   &   1.02        \\ \hline
    1000 &   2.62  &  1.06   \\ \hline
    50000&   2.15  &   1.08      \\ \hline
    
\end{tabular}
\end{table}

\begin{figure}[h!]
    \centering
    
    \begin{subfigure}{0.47\textwidth}
        \includegraphics[width=\linewidth]{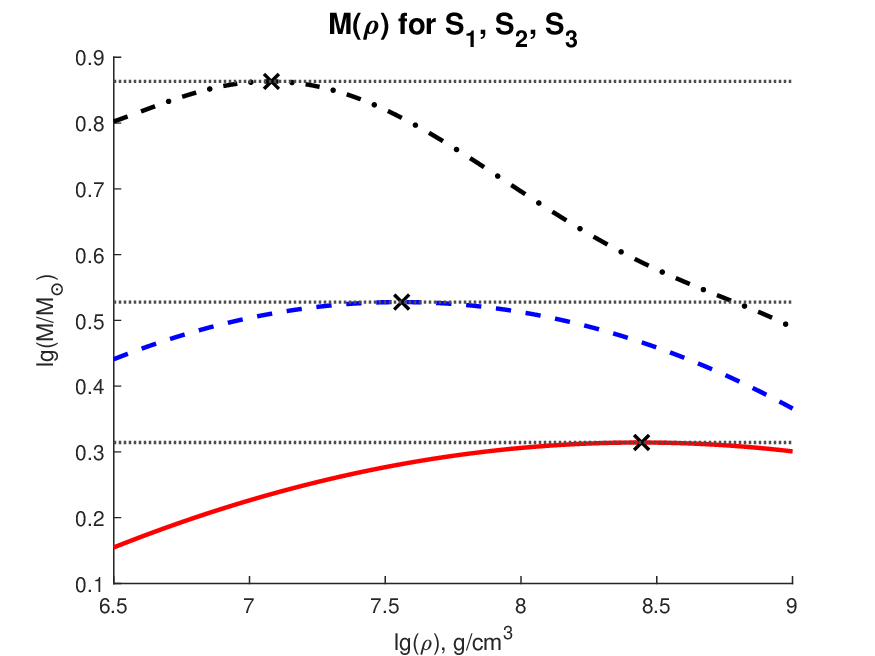}
       
    \end{subfigure}
    \hfill
    \begin{subfigure}{0.47\textwidth}
        \includegraphics[width=\linewidth]{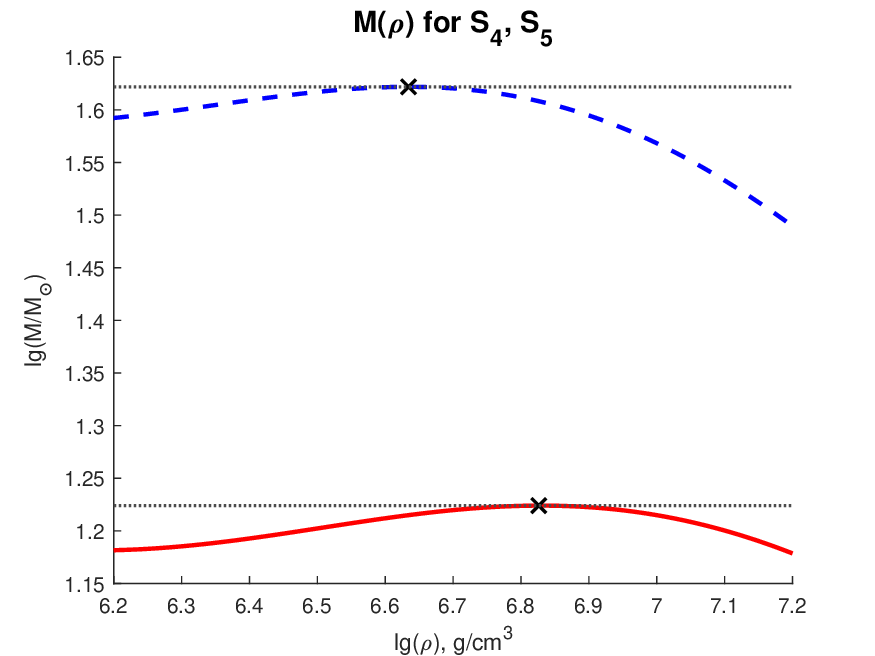}
       
    \end{subfigure}
    
    \begin{subfigure}{0.47\textwidth}
        \includegraphics[width=\linewidth]{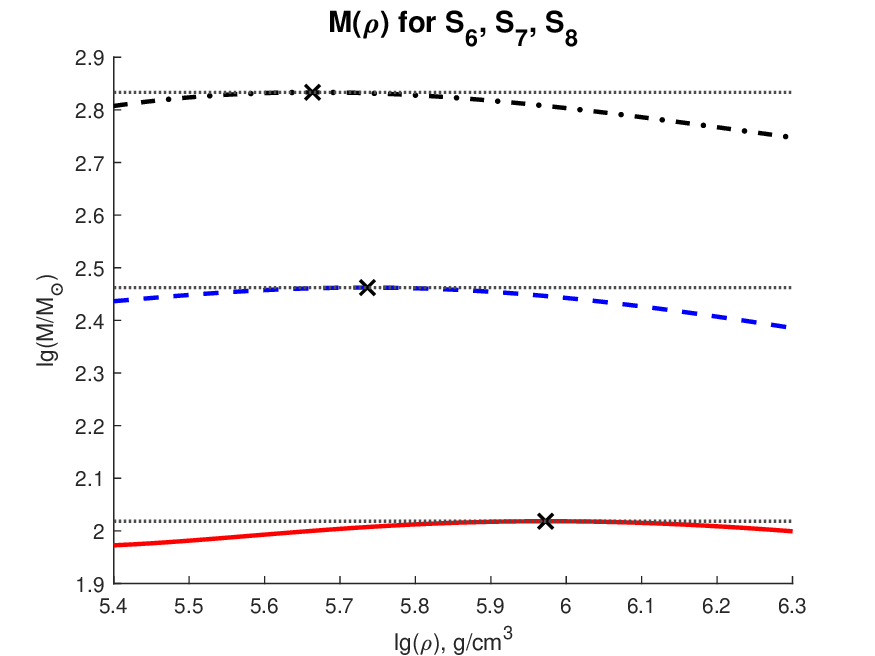}
        
    \end{subfigure}
    \hfill
    \begin{subfigure}{0.47\textwidth}
        \includegraphics[width=\linewidth]{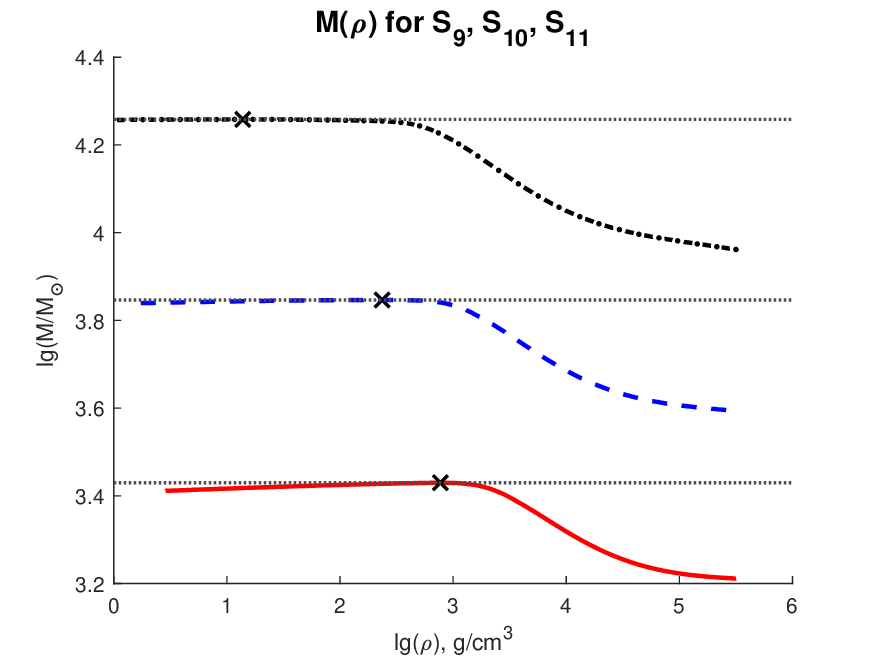}
        
    \end{subfigure}
    \\{{\bf Fig. 1 }  Dependency graphs $M(\rho)$ in the isentropic model of uniform density for different isentropes. Entropy is given 
in units of $10^{10}$ erg/K/g everywhere. Top left $-$ red solid line corresponds to isentrope $S_1 = 0.01$, blue dashed $-$ $S_2 = 0.01585$,  black dashed line $-$ $S_3 = 0.02512$. The black crosses mark the maxima corresponding to the critical state of the star. The dotted lines are
straight lines that pass through these maxima. Top right $-$ the same for isentropes: $S_4 = 0.03981, S_5 = 0.0631$. Bottom left $-$ the same for isentropes: $S_6 = 0.1, S_7 = 0.1585, S_8 = 0.2512$. Bottom right $-$ the same for isentropes: $S_9 = 0.3981, S_{10} = 0.631, S_{11} = 1$.}
\end{figure}

 \begin{figure}
     \centering
     \includegraphics[width=0.7\linewidth]{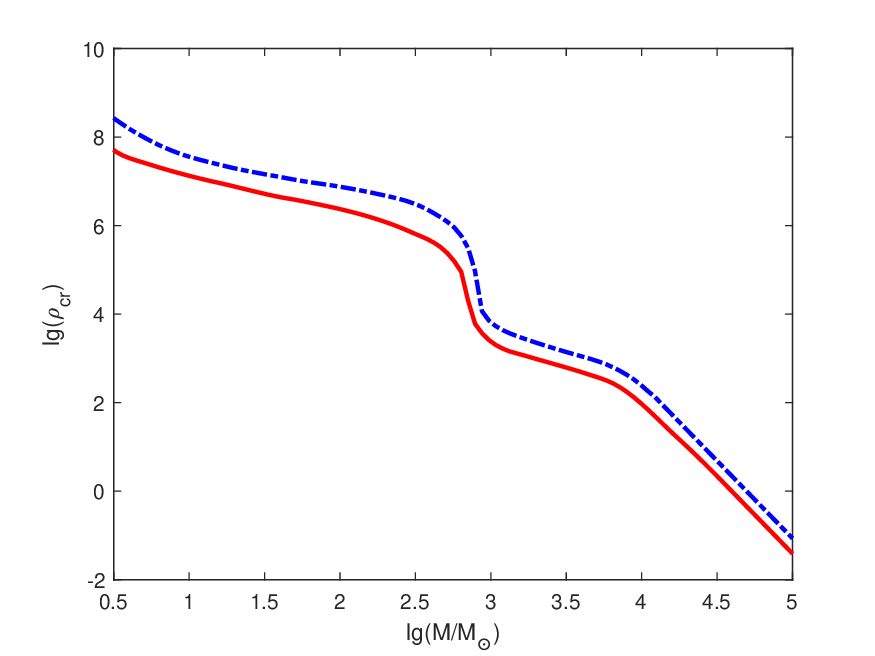}
     \\ {{\bf Fig.2} The dependence  $\rho_{cr}$(M) density in critical state from mass. The blue dashed line represents the critical central density \cite{bk}. The
red solid line corresponds to the uniform density model, constructed based on the results in Table 3.}
     \label{fig:enter-label}
 \end{figure}

  \begin{figure}
     \centering
     \includegraphics[width=0.7\linewidth]{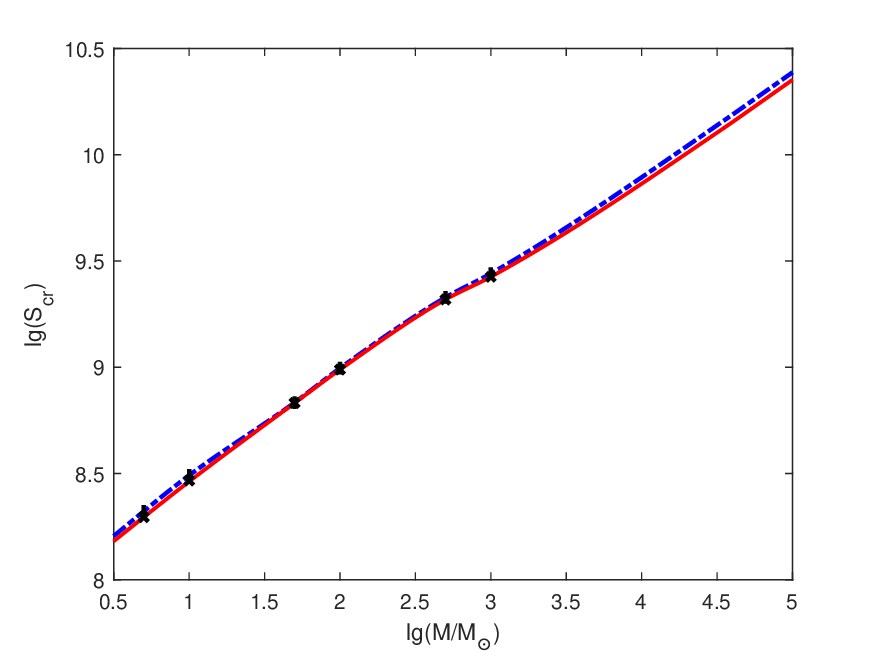}
     \\ {{\bf Fig.3} The dependence $S_{cr}$(M)  entropy in critical state from mass. The blue dashed line corresponds to the model from \cite{bk}. The red
solid line corresponds to the uniform density model, constructed based on the results in Table 3. The crosses correspond to the
models in the Table 4.}
     \label{fig:enter-label}
 \end{figure}

\section{Supermassive stars}

In an equilibrium star of large mass, the entropy is
so great that the pressure and energy of the matter are
determined mainly by radiation, and the share of gas
pressure is relatively small.\\

For pure radiation the adiabatic index is $\gamma = 4/3$, i.e., in Newtonian theory there is an indifferent equilibrium of the star at any density.  In massive stars the
adiabatic index is close to 4/3, so deviations $\gamma$ from 4/3  must be
carefully considered and  small corrections for general relativity should also be taken into
account. Deviations $\gamma$ from 4/3 is associated with the contribution to
the equation of state of the plasma pressure, due to
which $\gamma$  becomes greater than  $4/3$, and the star can be in stable hydrodynamic equilibrium. Due to the small
difference $\gamma$ from $4/3$ even small effects are enough to significantly affect the stability of the star.

We will act using the method of successive approximations. First, we find the equilibrium of the star
using Newtonian theory, taking into account only
radiation in energy and neglecting all corrections.
Then we will take into account the influence of plasma
and the role of general relativity. The processes of iron
dissociation and neutronization of matter are unimportant for an equilibrium supermassive star.

\subsection{Supermassive stars with radiation
pressure in Newtonian gravity }

The total energy of a star with Newtonian gravity in
the uniform density approximation is given in expression (1), where the last term responsible for the corrections for general relativity must be removed.

In the zeroth approximation, the radiation energy
per unit mass is considered $E_T$, entropy per unit
mass $S$, pressure $P$ along the isentropes,
at constant values $S$ and $K$, are determined by the
relations:
\begin{equation}
    E_{T} = \frac{a T^4}{\rho}, \quad S = \frac{4a T^3}{3\rho}, \quad P = \frac{1}{3}a T^4 = K\rho^{4/3}, \quad E_{T} = 3 K \rho^{1/3}, \quad
    \text{where} \quad S = \big (\frac{256 a}{3}\big)^{1/4} K^{3/4}.
    \label{eq:term_functions}
\end{equation}
Equilibrium of a star of a
given mass $M$ is determined by the extremum of the
total energy as a function of density $\rho$ at constant
entropy. It is convenient to write the derivative in
the form
\begin{equation}
    \frac{d\varepsilon}{d\rho^{1/3}} = 3 K M - \frac{3}{5}\big(\frac{4\pi}{3}\big)^{1/3}GM^{5/3} = 0.
    \label{eq:energy_derivative}
\end{equation}
The equilibrium in this case does not depend on the
density and is determined by a one-to-one relationship $K$ and $M$, following from expression (22), in
the form 
\begin{equation}
    K_{eq} = \frac{1}{5}\big(\frac{4\pi}{3}\big)^{1/3}GM^{2/3}= 0.3224 G M^{2/3}.
\end{equation}
From (21) and (22), we obtain the corresponding
value of the equilibrium entropy:
\begin{equation}
\label{eq8}
    S_{eq} = 4\sqrt 2 \left(\frac{\pi a}{9}\right)^{1/4} 
    \left(\frac{G}{5}\right)^{3/4}M^{1/2}= 
    7.1205\cdot10^7 \big ( \frac{M}{M_{\odot}} \big ) ^{1/2}.
\end{equation}
Temperature and density are related by the ratio
\begin{equation}
    T =\left(\frac{3K}{a}\right)^{1/4}\rho^{1/3} 
    =\left(\frac{3G}{5a}\right)^{1/4}
    \left(\frac{4\pi}{3}\right)^{1/12}
    M^{1/6}\rho^{1/3} =
    1.9183\cdot10^7 \big ( \frac{M}{M_{\odot}} \big ) ^{1/6} \rho^{1/3}
    \label{eq:T_rho}
\end{equation}

\subsection{Taking into account the influence of plasma}

When taking into account plasma in the form of
nuclei ($A,Z$) and electrons, the internal energy of the
matter is written as
\begin{equation}
    E = \frac{a T^4}{\rho} + \frac{3}{2}\frac{1+Z}{A m_p}kT
    \label{eq:E_plasma_T_rho}
\end{equation}
Let’s express the temperature $T$ through density $\rho$, and
total entropy $S$, considering the contribution of
plasma to be small. We have:

\begin{eqnarray}
        S = S_{rad} + S_{pl} = \frac{4a T^3}{3\rho} + \frac{k}{A m_p}\Big(\frac{5}{2} + \ln\big[ \big (\frac{A m_p k T}{2\pi\hbar^2} \big)^{3/2} \frac{g A m_p}{\rho}  \big] \Big) + \frac{Z k}{A m_p}\Big(\frac{5}{2} + \ln\big[ \big (\frac{m_e k T}{2\pi\hbar^2} \big)^{3/2} \frac{2 A m_p}{Z\rho}  \big] \Big), \nonumber \\
        \label{temp}
    T = \big(\frac{3 \rho}{4\sigma}\big)^{1/3}\Bigg(S - k\frac{1+Z}{Am_p} \ln\big(\frac{T^{3/2}}{\rho}\big)-{\rm Const}\Bigg)^{1/3}\qquad\qquad\qquad\qquad\qquad
\end{eqnarray}

\begin{equation}
\label{const}
{\rm Const}= \frac{k}{Am_p}
\Bigl[\frac{5}{2}(1+Z)+\ln\big[\big(\frac{A m_p k }{2\pi\hbar^2} \big)^{3/2} g A m_p\big]+ Z\ln\big[ \big(\frac{m_e k }{2\pi\hbar^2} \big)^{3/2} \frac{2 A m_p}{Z} \big] \Bigr] 
\end{equation}
Here,  $g$ - statistical weight of the core. When plasma is
a small additive, so that $\frac{S_{pl}}{S} \ll 1$, in the second term in
brackets in the expression for temperature (27), we use
the zero approximation for $T=T_0$  in the form of

\begin{equation}
T_0=\big(\frac{3 S \rho}{4a}\big)^{1/3}, 
\quad
     T = \big(\frac{3 S \rho}{4a }\big)^{1/3} \Bigg[1 - k\frac{1+Z}{3SAm_p}
   \ln(\frac{3 S}{4a\rho})^{1/2}
   -\frac{\rm Const}{3S}\Bigg].
    \label{eq:T_approach}
\end{equation}
Expression for internal energy (26) as a function $(\rho,S)$, has the form:

\begin{eqnarray}
  E(\rho,S) = \frac{a}{\rho}\big(\frac{3\rho S}{4a}\big)^{4/3} \Bigg[1 -4 k\frac{1+Z}{3SAm_p}
   \ln(\frac{3 S}{4a\rho})^{1/2}
   -4\frac{\rm Const}{3S}\Bigg]
  + \frac{3}{2}\frac{k(1+Z)}{Am_p}\big(\frac{3\rho S}{4a}\big)^{1/3}
       \nonumber \\
   = \frac{a}{\rho}\big(\frac{3\rho S}{4a}\big)^{4/3} 
   \Bigg[1 -\frac{4k}{3SAm_p}\Big\{
   \ln[(\frac{3 S}{4a\rho})^{1/2}(\frac{A m_p k }{2\pi\hbar^2} \big)^{3/2} g A m_p]
  +Z \ln[(\frac{3 S}{4a\rho})^{1/2}\big(\frac{m_e k }{2\pi\hbar^2} \big)^{3/2} \frac{2 A m_p}{Z}]
        \label{eq:E_pl} 
  \\ +\frac{10}{3}(1+Z)\Big\}\Bigg]  
   + \frac{3}{2}\frac{k(1+Z)}{Am_p}\big(\frac{3\rho S}{4a}\big)^{1/3}
   \nonumber
 \end{eqnarray}

\subsection{Equilibrium and stability of supermassive
stars in general relativity}
Total energy of a homogeneous star at a given
entropy $S$, taking into account the first corrections to
GTR is given in (1). The equilibrium of this star is
determined by the equality of the first derivative to
zero
  $ \frac{d \varepsilon}{d \rho} =0$ 
The critical state where the loss of stability occurs is determined by the equality of the second
derivative to zero
 $  \frac{d^2 \varepsilon}{d\rho^2} = 0$.
Instead of density derivatives $\rho$  it is more convenient to take derivatives with
respect to the value $\rho^{1/3}$.
 From (30), we have

\begin{eqnarray}
\frac{dE(\rho,S)}{d\rho^{1/3}} = a\big(\frac{3 S}{4a}\big)^{4/3} \Bigg[1 -4 k\frac{1+Z}{3SAm_p}
   \ln(\frac{3 S}{4a\rho})^{1/2}
   -4\frac{\rm Const}{3S}\Bigg]+\frac{3}{2}\frac{k(1+Z)}{Am_p}\big(\frac{3 S}{4a}\big)^{1/3} =
   \nonumber \\
 = \big(\frac{3 S}{4a}\big)^{1/3}\Bigg[\frac{3S}{4}+k\frac{1+Z}{Am_p}
  \Big(3-\frac{1}{2}\ln\frac{3S}{4a\rho}\Big)
  -{\rm Const}\Bigg]
  \end{eqnarray}       
The second derivative is defined as

\begin{equation}
    \frac{d^2 E(\rho,S)}{(d \rho^{1/3})^2} = \frac{3}{2}\frac{k(1+Z)}{A m_p}\Big(\frac{3S}{4a\rho}\Big)^{1/3}M 
\end{equation}
Using (2), (3), we write the equilibrium condition and
the equation for determining the critical point in the
form 

\begin{eqnarray}
  \label{eq10}
\frac{d\varepsilon}{d\rho^{1/3}}=\big(\frac{3 S}{4a}\big)^{1/3}\Bigg[\frac{3S}{4}+k\frac{1+Z}{Am_p}
  \Big(3-\frac{1}{2}\ln\frac{3S}{4a\rho}\Big)
  -{\rm Const}\Bigg]M -\qquad \\
 - \frac{3}{5}\big(\frac{4\pi}{3}\big)^{1/3}GM^{5/3} - 3.964\frac{G^2}{c^2}M^{7/3}\rho^{1/3}=0,\qquad
  \nonumber
\end{eqnarray}

\begin{equation}
    \label{eq11}
    \frac{d^2 \varepsilon}{(d\rho^{1/3})^2} = \frac{3}{2}\frac{k(1+Z)}{A m_p}\Big(\frac{3S}{4a\rho}\Big)^{1/3}M - 3.964\frac{G^2}{c^2}M^{7/3}=0.
\end{equation}
Connection between equilibrium entropy $S$ with mass, in the
zero approximation coinciding with (24), follows from
(33), if we neglect small terms. The critical density is
found from expressions (34), (24), after which the
critical temperature in the zero approximation is calculated using formula (29). The critical parameters of
supermassive stars in the uniform density approximation are determined by the relations ($\mu=\frac{A}{1+Z}$) 

\begin{align}
    S_{u,cr} = 7.1205\cdot10^7\Big(\frac{M}{M_{\odot}}\Big)^{1/2}, \label{eq:crit_formula1} \\
    \rho_{u,cr} = 1.13 \cdot 10^{17} \frac{1}{\mu^{3}} \Big(\frac{M}{M_{\odot}}\Big)^{-7/2} \frac{g}{cm^3}, \label{eq:crit_formula2}\\
     T_{u,cr} = 9.27 \cdot 10^{12} \frac{1}{\mu} \Big(\frac{M}{M_{\odot}}\Big)^{-1}K.
     \label{eq:crit_formula3}  
\end{align}
Due to the predominance of radiation pressure in an
isentropic equilibrium supermassive star, the density
distribution almost coincides with the polytropic one,
with $\gamma=\frac{4}{3}$.
For this model, the critical parameters of
supermassive stars, such as central density and temperature, entropy, were obtained in \cite{fowler,zeld2}, and are
determined by the formulas
\begin{align}
    S_{c,cr} = 7.72\cdot10^7\Big(\frac{M}{M_{\odot}}\Big)^{1/2}, \\
    \rho_{c,cr} = 2.43 \cdot 10^{17} \frac{1}{\mu^{3}} \Big(\frac{M}{M_{\odot}}\Big)^{-7/2} \frac{g}{cm^3}, \\
     T_{c,cr} = 1.23 \cdot 10^{13} \frac{1}{\mu} \Big(\frac{M}{M_{\odot}}\Big)^{-1} K
\end{align}
 Plotting the dependence of critical density on mass $\rho_{cr}(M)$ for supermassive stars in the homogeneous
model in Figs. 2 and 3 was done using formulas (35)$-$
(37) using $\mu = 56/27$ (iron). 

\section{Conclusions}

In our work, a tabular equation of state was
obtained for a number of isentropes in the range of
densities   $1<\rho< 10^{10}  \frac{\text{g}}{\text{cm}^3}$ and temperatures $10^8 < T < 10^{10} $ K. At high temperatures and densities,
where the relativistic degeneracy of electrons and positrons, the production of pairs and photons, $\beta -$processes and nuclear reactions in a
mixture {$^{56}Fe$},  {$^{4}He$}, $p, n$ are taken
into account, the results of work \cite{imsh} were
used. At lower densities and temperatures, the solution
of the system of equations (4) $-$ (7) was used. Using the
obtained equation of state, equilibrium curves $M(\rho)$ for various isentropes were
constructed, which made  it possible to determine the points of loss of stability of
stars in the range of masses $2 < \frac{M}{M_{\odot}} < 10^5$. All critical
parameters are listed in Table 3. Comparison of the
results of the uniform density model with the more
accurate model from \cite{bk} is presented in Table 4,
showed that the ratio of the central density in the more
accurate model to the uniform density, $\rho_{c, cr}/\rho_{u, cr}  $  is in
the range of 2.15 $-$ 11, and the entropy ratio is in the
range of 1.006 $-$ 1.12. The large difference in the ratio
of critical densities and the closeness of critical entropies for the same masses is due to the fact that in both
cases isentropic models were compared, and the density distributions in the compared models differed
greatly. In Fig. 2, for our model, the constant critical
density for the star is indicated vertically, and for the
model from work \cite{bk},  for lack of a better one, the central critical density is indicated, which should obviously be greater than the uniform one for the same
mass. The obtained weak sensitivity of the critical
entropy to a change in the profiling density used in
constructing the model can somewhat simplify the
finding of the parameters of the critical states of stars
of a given mass with a density distribution far from
polytropic. Isentropicity of a star in a critical state can
be expected if convection develops.

Using the results of this work to determine critical
parameters in models with uniform entropy makes this
procedure less labor-intensive. This is due to the fact
that the critical value of entropy can be found relatively
easily in the model of a homogeneous star. Using this
entropy value, one can then construct an accurate
model by solving the equilibrium equation of the star
with the same equation of state.

\bigskip
FUNDING

\bigskip

The work was supported by the Russian Science Foundation (Project No. 23-12-00198).


\end{document}